# New spin injection scheme based on spin gapless semiconductors: A first-principles study

G. Z. Xu[1], X. M. Zhang[1], Z. P. Hou[1,2], Y. Wang[1], E. K. Liu[1], X. K. Xi[1], S. G. Wang[1], W. Q. Wang[2], H. Z. Luo[3], W. H. Wang[1,a)], and G. H. Wu[1]

[1]*State Key Laboratory for Magnetism, Beijing National Laboratory for Condensed Matter Physics, Institute of Physics, Chinese Academy of Sciences, Beijing 100190, P. R. China*
[2] *College of Physics, Jilin University, Changchun 130023, P. R. China*
[3] *School of Material Science and Engineering, Hebei University of Technology, Tianjin 300130, P. R. China*

**Abstract** –Spin injection efficiency based on conventional and/or half-metallic ferromagnet/semiconductor is greatly limited by the Schmidt obstacle due to conductivity mismatch, here we proposed that by replacing the metallic injectors with spin gapless semiconductors can significantly reduce the conductive mismatch to enhance spin injection efficiency. By performing first principles calculations based on superlattice structure, we have studied the representative system of $Mn_2CoAl$/semiconductor spin injector scheme. The results showed that a high spin polarization was maintained at the interface in systems of $Mn_2CoAl/Fe_2VAl$ constructed with (100) interface and $Mn_2CoAl$/GaAs with (110) interface, and the latter is expected to possess long spin diffusion length. Inherited from the spin gapless feature of $Mn_2CoAl$, a pronounced dip was observed around the Fermi level in the majority-spin density-of-states in both systems, suggesting fast transport of the low-density carriers.

------------------------------

a) Author to whom correspondence should be addressed. Electronic mail: wenhong.wang@iphy.ac.cn

**Introduction. -** Rapid development of spintronics requires large sources of spin-polarized charge carriers, turning spin injection into a field of growing interest in recent decades. Conventionally, spin injection utilized the ferromagnet/semiconductor (SC) interface [1], for which the injection efficiency was greatly limited due to the conductivity mismatch (theoretically modeled by Schmidt *et al* [2]) and low spin polarization degree of the magnetic source. Subsequently, with the emerging of half-metallic ferromagnets (HMF) that possess nearly 100% spin polarization, the HMF/SC heterostructures were proposed for enhancing the spin injection efficiency [3, 4]. Nevertheless, the conductivity mismatch between the metal and semiconductor still exists. Tunnel contacts were raised as one way to circumvent this obstacle [5, 6]. In some cases, the metal and semiconductor interface can form natural Schottky barrier [7, 8], but more commonly a thin oxide layer was inserted to form tunneling barriers, such as FM/MgO/SC heterostrucutres [9, 10], where the spin injection efficiency has been greatly enhanced. On the other hand, magnetic semiconductors were also investigated to realize highly polarized spin current [11, 12], but they are restricted to low temperatures and sometimes need large field bias.

Here in this letter we proposed another spin injector scheme that not only keep a high spin polarization of the injection source like HMF, but also can effectively overcome the conductivity mismatch. The scheme we considered is based on the spin gapless semiconductors (SGS), a kind of gapless semiconductor accompanying with fully spin polarized charge carriers [13]. As seen in Fig. 1 for the DOS sketch of SGS, zero gap was realized in one spin direction (↑) like in gapless semiconductors, but in the other spin (↓) a finite gap was opened resulted from spin splitting, thus the transport carriers are theoretically fully spin polarized like in HMF. Owing to its expected semi-conductivity, it could be more compatible to the semiconducting substrate from the impedance.

Among the SGS candidates, Heusler alloy $Mn_2CoAl$ has been predicted to be a spin gapless semiconductor both theoretically and experimentally [14]. The reported data for the conductivity of $Mn_2CoAl$ is in the order of $10^3$ S cm$^{-1}$, about two orders lower than the traditional HMF (for example, $Co_2MnSi$ is ~$10^5$ S cm$^{-1}$ [15]). Considering that the electronic states of SGS are extremely sensitive to the atomic order, we evaluated the conductivity of fully ordered $Mn_2CoAl$ by employing BoltzTraP code [16] based on semiclassical Boltzmann transport theory. The calculated room temperature conductivity (with respect to a constant relaxation time) for the three typical systems of $Co_2MnSi$, $Mn_2CoAl$ and GaAs were presented in the right panel of Fig. 1. It can be seen that the conductivity of perfect $Mn_2CoAl$ is much lower (nearly ten times) than that of $Co_2MnSi$, while very close to that of GaAs. The small conductivity mismatch between SGS and SC is promising to enhance spin polarization degree in SC region according to the Schmidt model [2]. In addition, the Heusler type SGS we concerned in the present study possess the advantage of high Curie temperature and compatibility to the industrial semiconductor from both structure and lattice constant.

The interfacial spin polarization can be evaluated by building heterostructure model using first principles method, as done in most systems of HMF/SC [3, 17]. In the present letter we investigated the SGS/SC (SGS=$Mn_2CoAl$, SC=$Fe_2VAl$, GaAs) heterostructures in terms of its layer-by-layer properties. We found that high spin polarization can be preserved for certain interface configurations.

**Calculation details.**─The first-principle calculations were performed within the framework of density functional theory by pseudopotential method implemented in the CASTEP package [18, 19]. The exchange correlation energy was treated under the generalized gradient approximation (GGA) [20]. Theoretically, the spin injected system can be simulated with superlattice structure as long as the constitute layers are thicker enough to restore the bulk properties in the center part of the slabs[4]. In the present study, we investigated the heterostructure properties by constructing supercells containing several unit cells of the constituent material. Two typical kinds of semiconductor substrates were chosen: one is the non-magnetic Heusler alloy $Fe_2VAl$, which has experimentally been proved to exhibit semiconductor-like properties [21]; the other is the stereotype semiconductor of GaAs. In Fig. 2, the $Mn_2CoAl/Fe_2VAl$ constructed with (100) and $Mn_2CoAl/GaAs$ with (110) interface were presented separately. The in-plane lattice parameters (a, b) of the supercells were determined to be in accordance with the bulk substrate, and the parameter c along the stacking direction was manually optimized. Based on the experimental value of $a_{Fe2VAl}$ = 5.76 Å and $a_{GaAs}$ = 5.65 Å, the lattice constant set for the supercell is $a = b = a_{Fe_2VAl}/\sqrt{2}$ for the (100) geometry, and $a = a_{GaAs}$, $b = a_{GaAs}/\sqrt{2}$ for the (110) geometry. The lattice mismatch between $Mn_2CoAl$ (a=5.8 Å) and these two substrates are 0.7% and 2.7%, respectively. For all cases we apply a plane-wave basis set cut-off energy of 500 eV to ensure good convergence and a mesh of 12×12×4 k-points for (100) interface and 12×8×4 for (110) interface. All DOS curves were plotted with a smearing width of 0.05eV.

**$Mn_2CoAl/Fe_2VAl$ heterostructures.** — We first determine the effect of the interface on the structural and electronic properties of all-Heulser $Mn_2CoAl/Fe_2VAl$ heterostructures. As seen in Fig. 2, $Mn_2CoAl/Fe_2VAl$ heterostructures constructed along [001] direction consist of alternate Co-Mn, Mn-Al and Fe-Fe, V-Al atomic layers. The interfacial layers that combine the two compounds can be either Co-Mn/V-Al (Mn-Al/Fe-Fe, the other side) or Co-Mn/Fe-Fe (Mn-Al/V-Al, the other side). We calculated both of them and found that the superlattice with Co-Mn/Fe-Fe (Mn-Al/V-Al) interface exhibit no spin polarization at all. In all-Heusler situation, as also discussed in ref. [4], the final spin polarization of the heterostructures can be simply anticipated by examining the junction components. The so-called constructive junction layers are usually those being semiconducting or ferromagnetic that bridge their bulk neighbors, like Co-Mn/V-Al (Mn-Al/Fe-Fe) here. Therefore, we studied in detail for the superlattices of [$Mn_2CoAl/Fe_2VAl$]$_4$ with Co-Mn/V-Al (Mn-Al/Fe-Fe) interface. As shown in Fig. 3 for the [100] projected superlattice, 32 nonequivalent atoms are distributed in 16 atomic planes. To obtain the layer-resolved magnetic moment, we added the moment of each atom in one layer, and the local spin polarization was defined as $P = \frac{N_\uparrow(E_F) - N_\downarrow(E_F)}{N_\uparrow(E_F) + N_\downarrow(E_F)}$, where $N(E_F)$ is the density of states at the Fermi level. It should be pointed out that in experiments the definition of spin polarization can vary depending on the measuring methods [22], the form here is simplified but has been widely recognized as a theoretical estimate [23]. It can be seen in Fig. 3 that in the middle of each slab, the bulk moment was well reproduced, with the total moment to be 8.09$\mu_B$, deviating little from the bulk values of $Mn_2CoAl$ (2$\mu_B$ ×4). When approaching the interface (in the middle and boundary of the figure), the moment of the magnetic layer decreased and, on the other hand, a small moment was induced in the semiconductor layer. Correspondingly, the degree of spin polarization in $Mn_2CoAl$ side maintained high, but dropped quickly in the semiconductor side, implying a short spin diffusion length in this heterostructure.

The right panel of Fig. 3 shows detailed evolution of the layer-resolved DOS for four layers around the interface. In the minority spin, a large gap was observed for all layers. The overall DOS pattern showed small difference with layer change, indicating weak interface scattering resulted from high structure similarities of this system. Still, comparing the same composition layer, for example 6th and 8th ones (both Co-Mn layer), the DOS of the 8th layer revealed less pronounced exchange splitting due to interface bonding states. The spin splitting of the semiconducting layer (9th to 12th) reduced with increased distance from the interface, indicating by the symmetry change of the spin-resolved DOS. Notably, the Fermi level locate near to a valley in the majority spin, inherited from the gapless feature of SGS (DOS scheme in Fig.1), making it advantageous with lower carrier densities comparing to the traditional metallic injection source.

**$Mn_2CoAl$/GaAs heterostructures.** ─ To guide future experimental efforts, here, results for the $Mn_2CoAl$/GaAs heterostructures are also presented. For the GaAs substrate, the heterostructure constructed in [100] direction lost its polarization at the interface according to our calculations. Studies on $Mn_2CoAl$ (100) surface revealed that while Mn-Al terminated surface maintained the half-metallicity of the bulk, Co-Mn termination destroyed it [24]. In contact with Ga or As in our case, Mn-Al also lost its high spin polarization, with overall moment almost vanished. For the (110) interface, as seen in Fig. 4 ([110] projection of the lattice), each plane contains a full formula unit of the cubic phase, so they are expected to restore the properties of the bulk form. Consistent with our result, high spin polarization has been reported in other (110) connected full Heusler alloy and GaAs systems [17, 25]. There are also two kinds of atom-connected ways for the (110) interface, which can be denoted as Co-Ga (Al-As) or Co-As (Al-Ga) considering their bonding ways. As the atomic magnetic moments of the latter decreased much in the interface, we focus on the former condition in the succeeding discussion.

The magnetic moment and spin polarization with respect to different [$Mn_2CoAl$/GaAs]$_4$ planes were given in Fig. 4 with the same manner of Mn$_2$CoAl/Fe$_2$VAl. In the Mn$_2$CoAl side, the moment of each layer remained almost unchanged with the bulk value of 2$\mu_B$, while the corresponding spin polarization was largely reduced comparing with the bulk of 100%. From the DOS of the 3rd and 4th layer, small states emerged in the minority spin when getting near to the boundary, since the majority DOS is also very small, the spin polarization can be easily destroyed due to the compensation of the states at Fermi level. Prominently, on the other hand, high spin polarization was observed for all layers in the semiconducting side, which is most desirable in designing spin injector system. The DOS of the 5th and 6th still present strong exchange splitting, indicating probable long spin diffusion length in this system. Like in the case of Fe$_2$VAl, a pronounced dip was also found in the majority-spin DOS, suggesting low carrier concentration that may facilitate fast transport of electrons.

**Prospect.** —To realize the application of the above scheme, the first step is to fabricate well ordered SGS films on SC substrates. Recently, Mn$_2$CoAl films oriented in (100) direction was deposited on both silicon and GaAs substrates, exhibiting ferromagnetism and semiconducting-like transport properties [26, 27]. Future work will concentrate on enhancing the structure ordering and further controlling the film growth direction and terminated layers, which greatly affect the injected spin polarization degree according to above discussions.

**Summary.** —Using first-principles density functional calculations, we have investigated the spin injection in two representative system of Mn$_2$CoAl/SC (SC=Fe$_2$VAl, GaAs), based on the assumption that SGS/SC can reasonably enhance the spin injection efficiency by reducing the conductivity mismatch. The computed results showed that systems of Mn$_2$CoAl/Fe$_2$VAl constructed with (100) interface and Mn$_2$CoAl/GaAs with (110) interface were favored for maintaining high spin polarization. Particularly, in Mn$_2$CoAl/GaAs system, a high degree of spin polarization was achieved in the semiconducting region, implying a long spin diffusion length. Remarkably, in both systems, the layered DOS reveal the spin gapless feature with a dip in the majority spin, which means that the transport carriers should be relatively low. This may give rise to higher mobility of the carriers comparing to traditional metallic injection system.

**Acknowledgements**

This work is supported by the National Basic Research Program of China (973 Program 2012CB619405), National Natural Science Foundation of China (Grant Nos. 11474343 and 11571207), and China-Israel Joint Project (Grant No. 2013DFG13020).

# References

[1] P. R. Hammar, B. R. Bennett, M. J. Yang, and Mark Johnson, Phys. Rev. Lett. **83**, 203 (1999).

[2] G. Schmidt, D. Ferrand, L. W. Molenkamp, A. T. Filip and B. J. van Wees, Phys. Rev. B **62**, R4790 (2000).

[3] G. de Wijs and R. de Groot, Phys. Rev. B **64**, R020402(2001).

[4] S. Chadov, T. Graf, K. Chadova, X. Dai, F. Casper, G. Fecher, and C. Felser, Phys. Rev. Lett. **107**, 047202 (2011).

[5] E. I. Rashba, Phys. Rev. B **62**, R16267 (2000).

[6] A. Fert and H. Jaffrès, Phys. Rev. B **64**, 184420 (2001).

[7] H. J. Zhu, M. Ramsteiner, H. Kostial, M. Wassermeier, H.-P. Schönherr, and K. H. Ploog. Phys. Rev. Lett. **87**, 016601(2001).

[8] A. T. Hanbicki, B. T. Jonker, G. Itskos, G. Kioseoglou, and A. Petrou. Appl. Phys. Lett., **80**, 1240 (2002).

[9] X. Jiang, R. Wang, R. M. Shelby, R. M. Macfarlane, S. R. Bank, J. S. Harris and S. S. P. Parkin. Phys. Rev. Lett. **94**, 056601 (2005).

[10] S. H. Liang, et al., Phys. Rev. B **90**, 085310 (2014).

[11] R. Fiederling, M. Keim, G. Reuscher, W. Ossau, G. Schmidt, A. Waag and L. W. Molenkamp, Nature **402**, 787 (1999).

[12] Y. Ohno, D. K. Y.oung, B. Beschoten, F. Matsukura, H. Ohno, and D. D. Awschalom, Nature **402**, 790 (1999).

[13] X. L. Wang, Phys.Rev. Lett **100**, 156404 (2008).

[14] S. Ouardi, G. H. Fecher, and C. Felser, Phys.Rev. Lett. **110**, 100401 (2013).

[15] L. Ritchie, et al., Phys. Rev. B **68**, 104430 (2003).

[16] G. K. H. Madsen and D. J. Singh, Comput. Phys. Commun. **175**, 67 (2006).

[17] K. Nagao, Y. Miura, and M. Shirai, Phys. Rev. B **73**, 104447 (2006).

[18] M. C. Payne, M. P. Teter, D. C. Allan, T. A. Arias, and J. D. Joannopoolous, Rev. Mod. Phys. **64**, 1065 (1992).

[19] M. D. Segall, P. L. D. Lindan, M. J. Probert, C. J. Pickard, P. J. Hasnip, S. J. Clark, and M. C. Payne, J . Phys. : Condens. Mat. **14**, 2717 (2002).

[20] J. P. Perdew, K. Burke, and M. Ernzerhof, Phys. Rev. Lett. **77**, 3865 (1996).

[21] M. K. Y. Nishino, S. Asano, K. Soda, M. Hayasaki, and U. Mizutani, Phys. Rev. Lett. **79**, 1909 (1997).

[22] P A Dowben, N. Wu, and C. Binek, J. Phys.: Condens. Matter **23**, 171001 (2011).

[23] X. Kozina, J. Karel, S. Ouardi, S. Chadov, G. H. Fecher, C. Felser, G. Stryganyuk, B. Balke, T. Ishikawa, T. Uemura, M. Yamamoto, E. Ikenaga, S. Ueda, and K. Kobayashi. Phys. Rev. B **89**, 125116 (2014).

[24] J. Li and Y. Jin, Appl. Surf. Sci. **283**, 876 (2013).

[25] L.Y. Chen, S. F. Wang, Y. Zhang, J. M. Zhang, and K.-W. Xu, Thin Solid Films **519**, 4400 (2011).

[26] M. E. Jamer, B. A. Assaf, T. Devakul, and D. Heiman, Appl. Phys. Lett. **103**, 142403 (2013).

[27] G. Z. Xu, Y. Du, X. M. Zhang, H. G. Zhang, E. K. Liu, W. H. Wang, and G. H. Wu, Appl. Phys. Lett. **104**, 242408 (2014).

**Figure captions:**

**FIG. 1** (color online). Left: DOS scheme of the spin gapless semiconductor (up) and conventional semiconductor (down). Middle: a sketch of our calculation model of SGS/SC heterostructure. Right: the calculated room temperature conductivity with respect to a constant relaxation time ($\tau$) for $Co_2MnSi$, $Mn_2CoAl$ and GaAs, plotted as a function of chemical potential. The green bar centered on the Fermi level marks regime without doping.

**FIG. 2** (color online). The superlattice primitive cells of $[Mn_2CoAl/Fe_2VAl]_2$ with (100) interface (left) and $[Mn_2CoAl/GaAs]_4$ with (110) interface (right) used for the first principles calculations. Both of them contain eight atomic layers. The crystal lattice of $Mn_2CoAl$ and GaAs are given in the middle panel for better understanding of the stacking pattern.

**FIG. 3** (color online). Left panel: the structure of $[Mn_2CoAl/Fe_2VAl]_4$ superlattice projected in one [100] direction and the corresponding layer-resolved magnetic moment and spin polarization. The dotted line marks the junction position of the two compound layers. Right panel: the layer resolved DOS for the $Mn_2CoAl/Fe_2VAl$ interface. The number corresponds to the left indicated ones.

**FIG. 4** (color online). Left panel: the structure of $[Mn_2CoAl/GaAs]_4$ projected in [110] direction and the corresponding layer-resolved magnetic moment and spin polarization. Right panel: the layer resolved DOS for the $Mn_2CoAl/GaAs$ interface. The number corresponds to the left indicated ones.

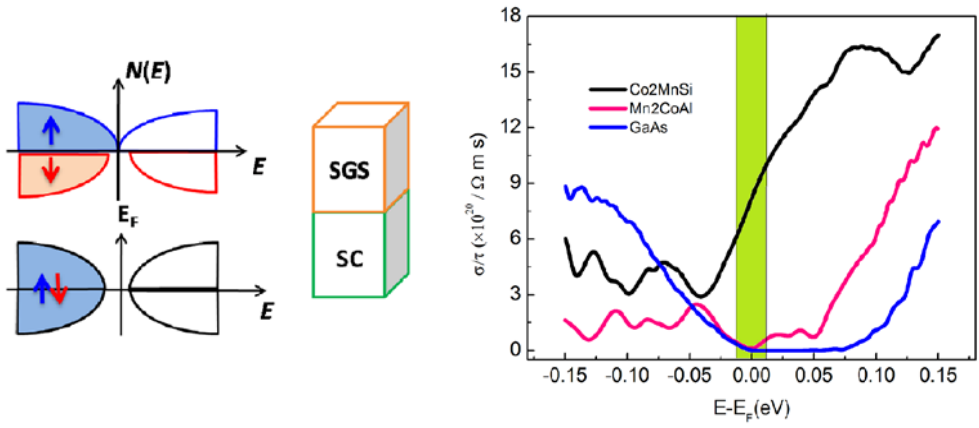

FIG. 1. Xu et al

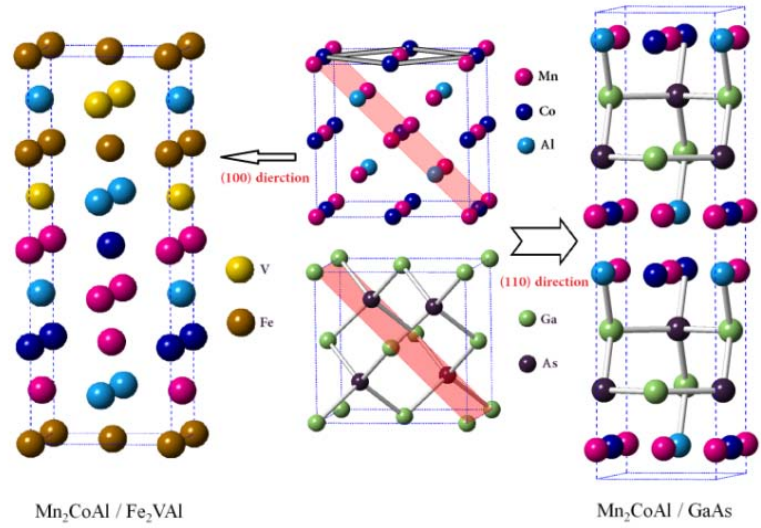

FIG. 2. Xu et al

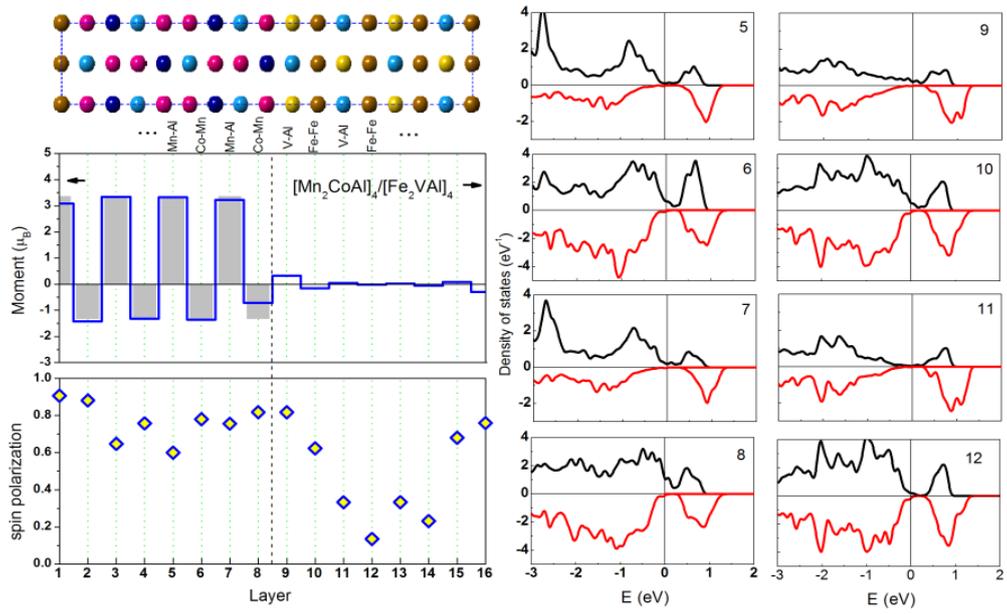

FIG. 3. Xu et al

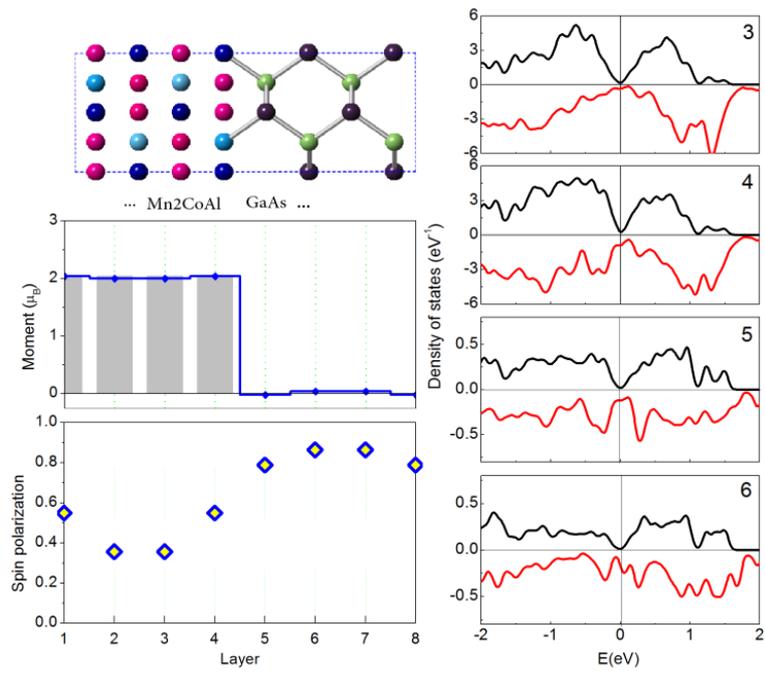

FIG. 4. Xu et al